\documentclass[twocolumn, twocolappendix, trackchanges]{aastex701}
\usepackage{graphicx}
\usepackage[normalem]{ulem}
\usepackage{dcolumn}   % Align table columns on decimal point
\usepackage{bm}        % Bold math
\usepackage[english]{babel}
\usepackage[dvipsnames]{xcolor}
\usepackage{hyperref}
\hypersetup{
    colorlinks = true,
    linkcolor  = Blue,
    urlcolor   = Blue,
    citecolor  = Blue
}
\usepackage{orcidlink}

\newcommand{\orcid}[1]{\href{https://orcid.org/#1}{\includegraphics[height=\fontcharht\font`\B]{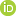}}}

\begin{document}

\title{Jet–environment interaction after delayed collapse in binary neutron star mergers}

\author[orcid=0000-0002-2945-1142,gname=Jay,sname=Kalinani]{Jay V. Kalinani}
\affiliation{Center for Computational Relativity and Gravitation \& School of Mathematical Sciences, Rochester Institute of Technology, 85 Lomb Memorial Drive, Rochester, New York 14623, USA}
\affiliation{The Grainger College of Engineering, Department of Physics \& Illinois Center for Advanced Studies of the Universe, University of Illinois Urbana-Champaign, Urbana, Illinois 61801, USA}
\email{jaykalinani@gmail.com}

\author[orcid=0000-0003-3140-8933,gname=Riccardo,sname=Ciolfi]{Riccardo Ciolfi}
\affiliation{INAF, Osservatorio Astronomico di Padova, Vicolo dell’Osservatorio 5, I-35122 Padova, Italy}
\affiliation{INFN, Sezione di Padova, Via Francesco Marzolo 8, I-35131 Padova, Italy}
\email{riccardo.ciolfi@inaf.it}

\author[orcid=0000-0002-8659-6591,gname=Manuela,sname=Campanelli]{Manuela Campanelli}
\affiliation{Center for Computational Relativity and Gravitation \& School of Mathematical Sciences, Rochester Institute of Technology, 85 Lomb Memorial Drive, Rochester, New York 14623, USA}
\email{mxcsma@rit.edu}

\author[orcid=0000-0002-6947-4023,gname=Bruno,sname=Giacomazzo]{Bruno Giacomazzo}
\affiliation{Dipartimento di Fisica G. Occhialini, Università di Milano-Bicocca, Piazza della Scienza 3, I-20126 Milano, Italy}
\affiliation{INFN, Sezione di Milano-Bicocca, Piazza della Scienza 3, I-20126 Milano, Italy}
\email{bruno.giacomazzo@unimib.it}

\author[orcid=0000-0001-9476-1270,gname=Andrea,sname=Pavan]{Andrea Pavan}
\affiliation{INAF, Osservatorio Astronomico di Padova, Vicolo dell’Osservatorio 5, I-35122 Padova, Italy}
\affiliation{INFN, Sezione di Padova, Via Francesco Marzolo 8, I-35131 Padova, Italy}
\email{andrea.pavan@inaf.it}

\author[orcid=0000-0001-9505-6557,gname=Allen,sname=Wen]{Allen Wen}
\affiliation{Center for Computational Relativity and Gravitation \& School of Mathematical Sciences, Rochester Institute of Technology, 85 Lomb Memorial Drive, Rochester, New York 14623, USA}
\email{acw6923@g.rit.edu}

\author[orcid=0000-0002-7541-6612,gname=Yosef,sname=Zlochower]{Yosef Zlochower}
\affiliation{Center for Computational Relativity and Gravitation \& School of Mathematical Sciences, Rochester Institute of Technology, 85 Lomb Memorial Drive, Rochester, New York 14623, USA}
\email{yrzsma@rit.edu }

\begin{abstract}

We present general relativistic magnetohydrodynamic simulations of binary neutron star (BNS) mergers, where the collapse of the metastable massive neutron star (MNS) remnant leads to the production of an incipient jet having terminal Lorentz factor and Poynting-flux luminosity compatible with a short gamma-ray burst (GRB). 
We consider different MNS lifetimes of about 25 and 50\,ms, long enough for massive polar outflows to emerge before black hole (BH) formation. The interaction of the following BH-driven jet with such polar outflows, responsible for shock heating and possible electromagnetic signatures, is self-consistently captured for the first time. Exploiting an unprecedentedly low numerical density floor scaling as $r^{-6}$, we explore the jet propagation up to distances of $\sim\!10^4$\,km. Comparing the outcome of different MNS lifetimes, we find that the latter, by strongly affecting the propagation environment, plays a major role in determining the final properties of the escaping jet. Finally, we consider a non-collapsing case, where the MNS-driven outflow is found to exhibit a much higher density and lower velocity compared to the BH-driven jet.

\end{abstract}



% =====================
% Introduction
% =====================
\section{Introduction}

The coincident detection of gravitational waves (GWs) and electromagnetic (EM) radiation from the binary neutron star (BNS) merger event GW170817 \citep{abbott2017gw170817, abbott2017multi, pian2017spectroscopic, abbott2017grb, mooley2018superluminal, lazzati2018late} conclusively demonstrated that BNS mergers can power short gamma-ray burst (GRB) jets. However, significant uncertainties remain regarding the physical conditions under which these relativistic jets can form and successfully produce a short GRB.

General relativistic magnetohydrodynamic (GRMHD) simulations currently represent the essential tool to self-consistently investigate these physical conditions. Previous GRMHD studies have demonstrated that black-hole (BH)-powered incipient jets with promising GRB-compatible properties at their base, including terminal Lorentz factors $\Gamma_\infty\!\gtrsim\!100$, can emerge in cases where the massive neutron star (MNS) resulting from the merger collapses within relatively short timescales (up to a few tens of ms; e.g., \citealt{Ruiz2016, Ruiz2021, sun2022jet, Bamber2025}) \footnote{We also note the recent work of \cite{hayashi2024jet} considering the prompt collapse scenario, finding jets with $\Gamma_\infty \sim 10$ that are likely insufficient to power typical short GRBs.}.
Nevertheless, realistic MNS remnants in nature may collapse to a BH over a much broader range of lifetimes and still allow for jet formation, as suggested by several observational constraints (e.g., \citealt{Siegel2019}). Prior to collapse, longer-lived remnants can launch massive polar outflows \citep{Ciolfi2020, Moesta2020, combi2023jets, most2023impact, Kiuchi2024, bamber2024jetlike, aguilera2024delayed, musolino2024impact}, which then act as critical obstacles for the propagation of a subsequent BH-driven jet (e.g., \citealt{Pavan2025} and refs.~therein). 
The presence of such pre-collapse outflows may also be potentially responsible for distinctive, relatively unexplored EM signals preceding the main GRB emission (possibly connected to observed GRB precursors; \citealt{Troja2010, Minaev2017, zhong2019precursors}). 

In this work, we perform GRMHD simulations of BNS mergers where the MNS remnant collapses to a BH either $\simeq 25$ or $\simeq 50$\,ms post-merger, after which a GRB-compatible incipient jet is produced.
Such lifetimes are sufficient to produce substantial pre-collapse outflows, whose self-consistent interaction with the incoming BH-driven jet is shown here for the first time. Moreover, the two different collapse times are found responsible for major differences in the incipient jet propagation, demonstrating that the MNS lifetime is a central parameter when considering GRB production in BNS mergers.

Additionally, we explore a case where no collapse occurs. In line with previous works \citep{Ciolfi2020, Moesta2020, combi2023jets, most2023impact, Kiuchi2024, bamber2024jetlike, aguilera2024delayed, musolino2024impact}, we find the natural emergence of steady, collimated outflows, which remain however much denser and slower than the BH-driven jets produced in the other cases.

On a technical side, we employ an unprecedentedly low numerical density floor, avoiding any artificial effect of the latter on the relevant outflows, even at increasingly large scales. This represents a key requirement for reliably connecting numerical models to the observed GRB phenomenology.

% =====================
% Methodology
% =====================
\section{Physical Models and Numerical Setup}

We consider an equal-mass BNS system with individual gravitational masses of $1.3624\,M_\odot$, corresponding to a chirp mass of $1.186\,M_\odot$ (the same as the GW170817 event; \citealt{Abbott2019}).
The initial data are generated with the {\tt LORENE} library, with coordinate separation of 45\,km and adopting a piecewise-polytropic approximation of the APR4 equation of state (EOS), as implemented in \cite{Endrizzi2016}. 
For the evolution, we consider a hybrid EOS obtained by adding a thermal component to the `cold' piecewise-polytropic part, with the resulting pressure dependence on rest-mass density and specific internal energy given as
\begin{equation} \label{hybridEOS}
{
P(\rho, \epsilon) = P_{\rm cold} (\rho) + (\gamma_{\rm th} -1)(\epsilon - \epsilon_{\rm cold}(\rho)) \rho \,.
}
\end{equation}
For rest-mass densities above $10^{10}$\,g/cm$^3$, we adopt the widely used adiabatic index $\gamma_{\rm th}\!=\!1.8$ (e.g.~\citealt{Endrizzi2016} and refs.~therein), while for $\rho\!<\!10^{10}$\,g/cm$^3$ we set $\gamma_{\rm th}\!=\! 4/3$.  
The latter choice is motivated by the fact that realistic temperature-dependent EOS have an effective ideal-gas adiabatic index of 4/3 at those low densities (\citealt{Bauswein2010}; see Appendix \ref{stepfunc}).

Dipolar magnetic fields are initialized onto the two NSs at two orbits prior to merger, using the vector potential prescription of \cite{Moesta2020} with fields also extending outside the NS surface. The field strength at the NS poles is set as $B_{\rm pole}\!\simeq\!10^{15}$\,G, corresponding to a total magnetic energy of $E_{\rm mag} \simeq 3 \times 10^{49}$\,erg at merger time. 
While these pre-merger magnetic fields are unrealistically strong, they still represent a small perturbation. 
Such field strengths are chosen to achieve the high magnetization levels expected in the post-merger phase, despite the fact that the corresponding amplification mechanisms are not fully resolved. This is a common approach in studies presenting jet-forming BNS merger simulations (e.g.~\citealt{Ruiz2021,Bamber2025}), which are all limited by insufficient resolution. 

Our simulations are performed using the GRMHD code \texttt{Spritz} \citep{Cipolletta2020, cipolletta2021}, employing the generalised Lorenz gauge to evolve the vector potential 
\citep{Farris2012}, PPM reconstruction along with HLLE Riemann solver, RK4 method for time-stepping, and the publicly available \texttt{RePrimAnd} library for primitive variable recovery \citep{Kastaun2021, Kalinani2022}.
Spacetime evolution follows the BSSN formulation as implemented in the \texttt{MacLachlan} module of the Einstein Toolkit \citep{etk2023}.
Our Cartesian grid setup includes twelve refinement levels, with the two finest levels tracking each NS during the inspiral phase. 
At merger, we transition to fixed mesh refinement, and upon collapse to a BH (if any) we cover the latter with a cubic shaped extra level with $\simeq\!9.5$\,km side. The computational domain extends up to 51,500\,km in positive and negative $x,y,z$ directions.

A key novelty compared to previous work concerns the setting of the artificial atmosphere, i.e.~the minimum rest-mass density $\rho_\mathrm{atm}$ allowed during evolution. 
We recall that in \texttt{Spritz}, whenever the evolution would lead to $\rho\!<\!\rho_\mathrm{atm}$, density is reset to $\rho_\mathrm{atm}$, pressure is reset according to the cold part of the EOS, fluid velocity is set to zero, and magnetic fields remain unchanged. 
Here, we set $\rho_\mathrm{atm}\!=\!\rho^*\!=\!6.17\times10^{4}$\,g/cm$^3$ up to a radial distance of $r^*\!=\!74$\,km, while at larger distances we employ the power-law profile $\rho_\mathrm{atm}\!=\!\rho^*\,(r/r^*)^{-6}$. 
With this setting, no atmospheric floor reset occurs at any time above $\approx\!90$\,km distance, leaving all outflows, from dynamical ejecta to the incipient jet, completely unaffected. 
The $r^{-6}$ falloff is chosen following \cite{Pavan2021,Pavan2023}, which demonstrated that such a steep decline is necessary to guarantee a reliable large-scale evolution. A shallower prescription like, e.g., the one employed in \cite{hayashi2024jet} ($\rho_\mathrm{atm} \!\propto\!r^{-3}$ down to 0.166\,g/cm$^3$ and then constant) yields atmosphere densities orders of magnitude higher than ours beyond $700$\,km distance and $\sim\!1000$ times higher at 2000\,km, eventually becoming a critical limitation at increasingly large scales (see Appendix \ref{floorpres}).

We present four different simulations, i.e.~cases A-D.
In case A, in order to explore the scenario in which a BH-disk system powers an incipient jet, we manually induce the collapse of the NS remnant at $\approx\!25$\,ms after merger, employing the procedure introduced in \cite{Ciolfi2019}. 
In case B, we consider the non-collapsing scenario. 
In both cases, the finest grid spacing is set to $dx\!=\!177$\,m ($88.5$\,m in the extra level covering the BH in case A). 
Case C is a lower resolution version of case A, with finest spacing $dx\!=\!200$\,m ($100$\,m around the BH).
Case D is the same as case C except that the collapse is induced around 50\,ms after merger.
All cases are evolved up to 75\,ms after merger, except for case C, reaching 160\,ms after merger.

% =====================
% Results
% =====================
\section{Results} 
\label{results} 

In all models, the merger produces outflows of unbound material (i.e.~dynamical ejecta) in the form of equatorial tidal tails and the more isotropic shock-driven ejecta. 
The latter, which may affect at later times the propagation and observational signatures of an emerging jet, are characterized by a very fast low-density tail reaching velocities of $\approx\!0.9\,c$. 
Notably, following dynamical ejecta up to their natural front is made possible by the very low and decaying atmospheric floor (see Appendix \ref{floorpres}). 
At the same time, we note that, at present resolution, the dynamics of the fastest ejecta layer may be influenced by numerical effects, especially at large distances (see \citealt{Rosswog2025} for an in-depth analysis).

The merger results in a differentially rotating MNS. At 20\,ms post-merger, a helical magnetic tower has formed, extending to over 500\,km along the NS spin axis, where magnetic field strength is $\sim\!10^{14}$\,G.
This tower originates polar outflows with velocities up to $\simeq\!0.2\,c$. 
The collapse to a BH, $\simeq\!25$\,ms after merger in cases A, C and $\simeq\!50$\,ms after merger in case D, hampers the further development of such outflows and leads to a very different evolution with respect to case B, in which no collapse occurs. 
\begin{table}[t]
\setlength{\tabcolsep}{5pt}
\renewcommand{\arraystretch}{1.15}
\caption{Black hole (BH) and disk properties for cases A, C, and D, shown at 1\,ms and at 25\,ms after collapse.}
\label{tab:bh_disk_props}
\begin{ruledtabular}
\begin{tabular}{lccccccc}
& \multicolumn{3}{l}{1\,ms after collapse}
& \multicolumn{4}{l}{25\,ms after collapse} \\
Case & $M_{\rm BH}$ & $\chi$ & $M_{\rm disk}$ & $M_{\rm BH}$ & $\chi$ & $M_{\rm disk}$ & $\dot M$ \\
& $[M_\odot]$ & & $[M_\odot]$ & $[M_\odot]$ & & $[M_\odot]$ & $[M_\odot\,{\rm s}^{-1}]$ \\
\hline
A, C & 2.46 & 0.62 & 0.17 & 2.52 & 0.64 & 0.11 & 1.0 \\
D & 2.42 & 0.58 & 0.2 & 2.45 & 0.59 & 0.16 & 0.9 \\
\end{tabular}
\end{ruledtabular}
\end{table}

The BH and disk properties of collapsing models A and D are summarized in Table~\ref{tab:bh_disk_props}, where we report values at 1\,ms and at 25\,ms after collapse. Results for case A apply also to case C. For a later collapse, less material is directly swallowed by the BH, leaving a more massive disk. 

Right after collapse, as matter close to the BH is accreted, polar funnels are rapidly excavated along the spin axis. However, magnetic flux within the helical magnetic tower is in good part preserved and, in a few ms, radial velocities in the polar regions and above $\sim\!100$\,km distance start to reverse. This leads to a new bipolar highly collimated outflow developing along the BH spin axis, which then matures into a magnetically dominated incipient jet around 15\,ms after collapse.

\begin{figure} 
\includegraphics[width=1\columnwidth]{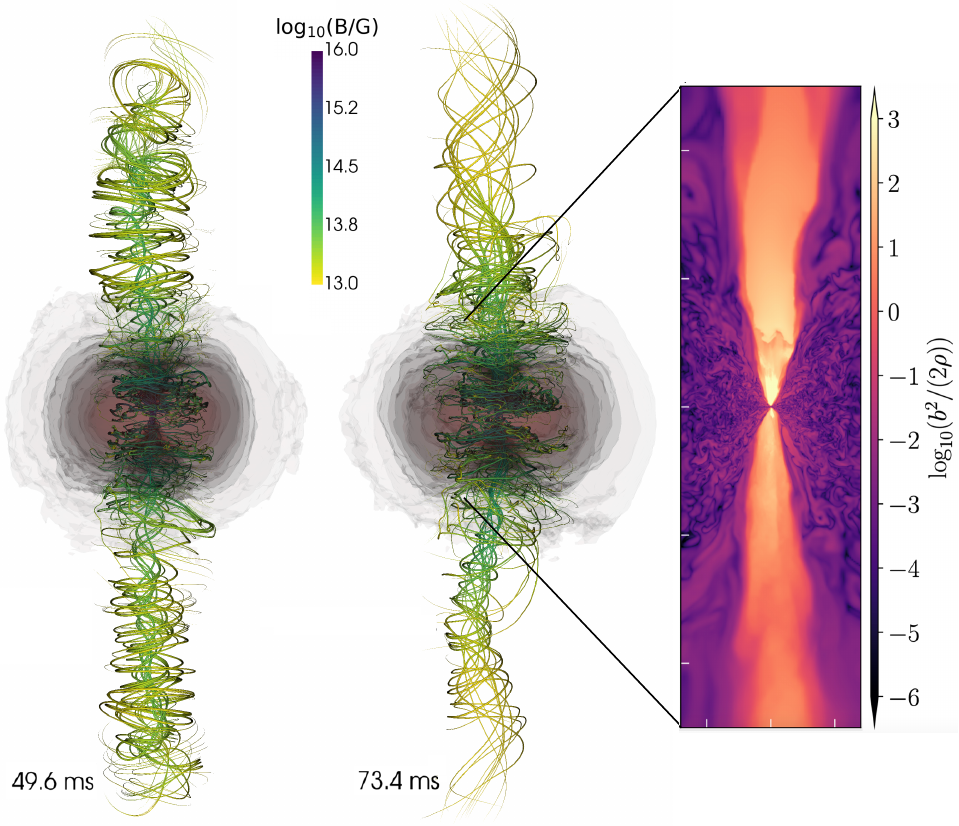}
\caption{3D rendering of magnetic field-line structure and rest-mass density (red-to-grey colors) for case A, at about $50$\,ms (left) and 75\,ms (right) after merger.
The inset shows a meridional view of the magnetization $b^2/(2\rho)$ at the jet base (vertical scale of up to $\pm 250$\,km).}
\label{fig:3D}
\end{figure}

In Figure \ref{fig:3D}, we show the magnetic field configuration along with rest-mass density distribution for case A, at about 25 and 50\,ms after BH formation. 
At both times, the average magnetic field strength around 500\,km distance along the spin axis is $\sim\!10^{14}$\,G, while the (mostly toroidal) field close to the BH and in the accretion disk has a strength in the range $10^{15}-10^{16}$\,G.

The inset of Fig.~\ref{fig:3D} gives a meridional view of the magnetization level, in terms of the magnetic energy density to rest-mass density ratio $b^2/(2\rho)$. When $b^2/(2\rho)\!\gg \!1$ at the base of the jet, this quantity gives an estimate of the terminal Lorentz factor $\Gamma_\infty$. 
Close to the BH, values $>\!10^3$ are achieved both for case A and case C, while $\Gamma_\infty$ reaches a few hundred in case D (see Appendix \ref{colltime}).
A GRB-compatible jet has to reach Lorentz factors of at least a few tens, if not $\gtrsim\!100$, which seems to be possible, at least in principle, for the BH-driven jets considered here. 
We caution, however, that $\Gamma_\infty$ does not give a prediction on the final Lorentz factor, but only an upper limit on what could be possibly achieved assuming full energy conversion into kinetic form and excluding any dissipation.
In other words, high enough values at the jet's base represent a necessary but not sufficient condition to be consistent with a GRB (see further discussion below).

\begin{figure} 
\includegraphics[width=1\columnwidth]{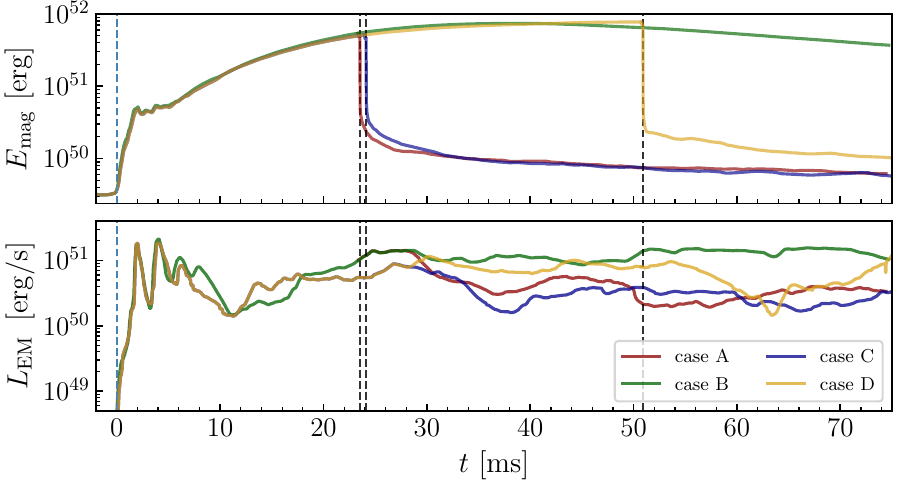}  
\caption{Evolution of total magnetic energy (top) and Poynting-flux luminosity across the $r\!=\!295$\,km spherical surface (bottom), for all cases. The vertical light blue line marks the merger time, whereas vertical black lines mark the collapse time for case A, C, and D, respectively.}
\label{fig:emag}
\end{figure}

Figure \ref{fig:emag} (top-panel) shows the time evolution of the total magnetic energy for the four cases. 
The initial exponential growth associated with the Kelvin-Helmholtz (KH) instability is followed by further amplification sustained by the massive NS differential rotation, via magnetic winding and the magnetorotational instability (MRI) (e.g., \citealt{Ciolfi2020b} and refs.~therein). 
We caution, however, that our resolution is insufficient to fully capture the KH instability as well as the MRI (and the possible $\alpha\Omega$ dynamo driven by it; \citealt{Kiuchi2024}).
While in case B magnetic energy keeps growing up to almost $10^{52}$\,erg and then starts to slowly decrease, in cases A and C the collapse leads to a sudden drop by a factor of $\approx\!60$, after which a slowly decreasing trend leads to a final value of $\simeq\!6\!\times\!10^{49}$\,erg. Case D indicates that a longer remnant lifetime enhances the outflow of magnetized material into the environment, yielding nearly twice the magnetic energy at the final time compared to cases A and C.

Bottom panel of Figure \ref{fig:emag} presents the evolution of the Poynting-flux luminosity $L_{\rm EM}$ across the spherical surface at $r\!=\!295$\,km. Here, the first few peaks within 8\,ms after merger relate to the dynamical ejecta crossing the surface.
Subsequently, the developing magnetic tower starts to drive a magnetized outflow, gradually increasing $L_{\rm EM}$. 
For case B, $L_{\rm EM}$ reaches $\sim\!10^{51}$\,erg/s around 25\,ms after merger, and this level is maintained until the end. 
In cases A, C and D, the collapse triggers a temporary drop until $\approx\!15$\,ms after BH formation, after which the contribution of the incipient jet becomes apparent. In the following evolution, $L_{\rm EM}$ remains rather stationary for cases A and C, settling at $\simeq 3\times10^{50}$\,erg/s, while for case D it increases to $\simeq 10^{51}$\,erg/s.  
These values are comparable to the power expected from the Blandford-Znajek mechanism \citep{Blandford1977,Thorne1986}, 
\begin{equation} 
{
L_{\rm BZ} \sim 5\!\times\!10^{50} \Bigg( \frac{\chi}{0.65} \Bigg)^{\!2} \Bigg( \frac{M_{\rm BH}}{2.4\,M_{\odot}} \Bigg)^{\!2} \Bigg( \frac{B_{\rm BH}}{5\!\times\!10^{15}\, {\rm G}} \Bigg)^{\!2} \, {\rm erg/s} \, .
}
\nonumber
\end{equation}

\begin{figure} 
\begin{center}
\includegraphics[width=1\columnwidth]{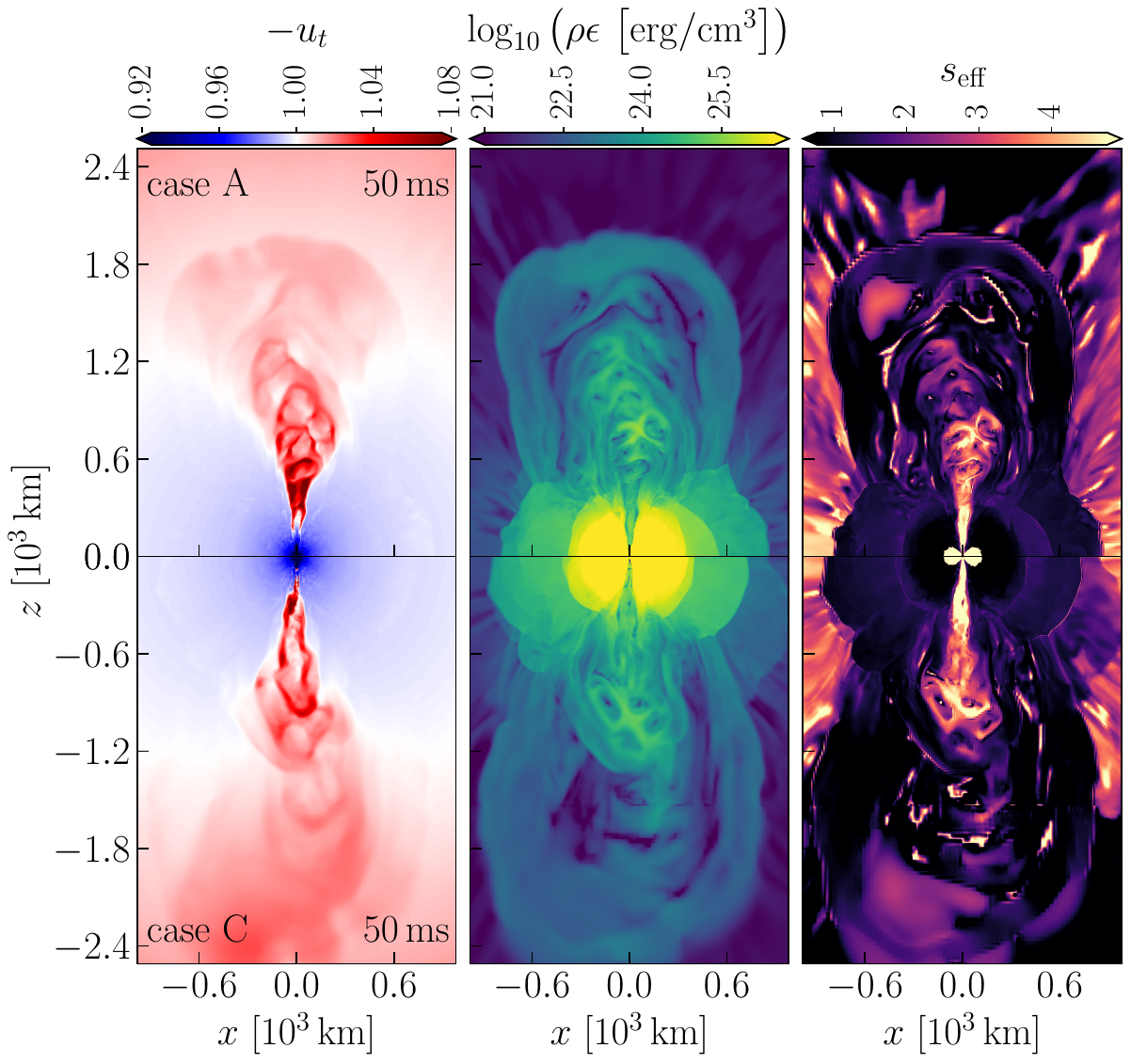}  
\caption{Meridional view of $-u_t$ (left), internal energy density $\rho\epsilon$ (center), and effective specific entropy $s_{\rm eff}$ (right) at $t\!\simeq\!50$\,ms after merger
for case A ($z\!>\!0$, top) and C ($z\!<\!0$, bottom).}
\label{fig:utes}
\end{center}
\end{figure}

Figure~\ref{fig:utes} compares cases A (top) and C (bottom) at $t\!\simeq\!50$\,ms after merger (i.e.~$t\!-\!t_{\rm BH}\!\simeq\!25$\,ms), showing unbound matter ($-u_t\!>\!1$, according to geodesic criterion), internal energy density $\rho\epsilon$, and effective specific entropy $s_{\rm eff}\!\equiv\! \ln(P/ \rho^{\gamma_{\rm eff}})$ computed from the effective adiabatic index $\gamma_{\rm eff}\!\equiv\!P/(\rho \epsilon) + 1$. In both collapsing scenarios, the incipient jet launched by the BH–disk system is highly unbound, and its front eventually encounters the more dense polar outflow expelled during the MNS phase. This interaction leads to alternating phases of jet advancement and stalling and produces shock heating, clearly visible as localized enhancements in $\rho\epsilon$ and sharp entropy jumps around $10^3$\,km distance
and $\lesssim\!15^\circ$ from the $z$-axis. 
These shocks, while dissipating jet energy, may leave a detectable imprint in the following EM emission, possibly powering a GRB precursor signal. Moreover, by modifying the thermodynamic conditions of the involved material, they can influence nucleosynthesis and possibly affect the resulting kilonova emission.

\begin{figure} 
\begin{center}
\includegraphics[width=0.555\columnwidth]
{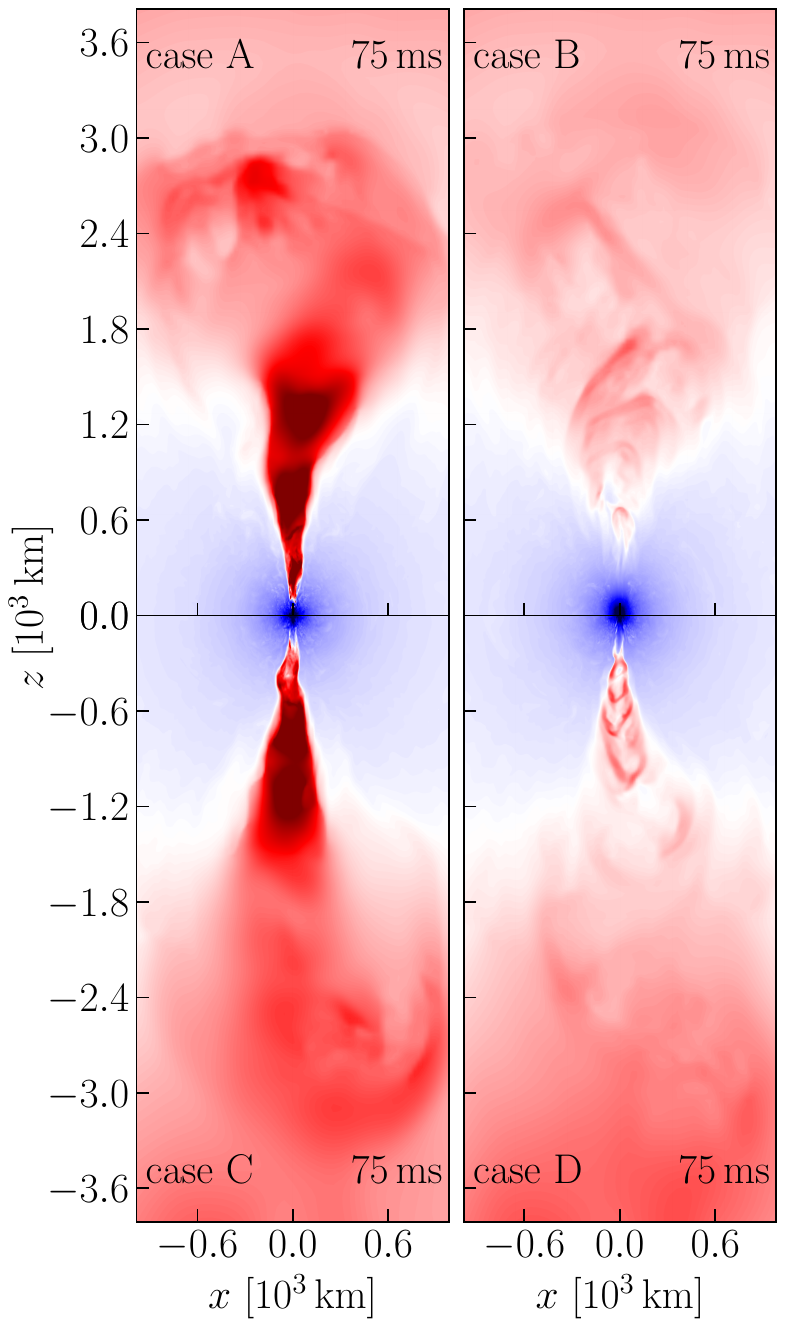} 
\includegraphics[width=0.426\columnwidth]
{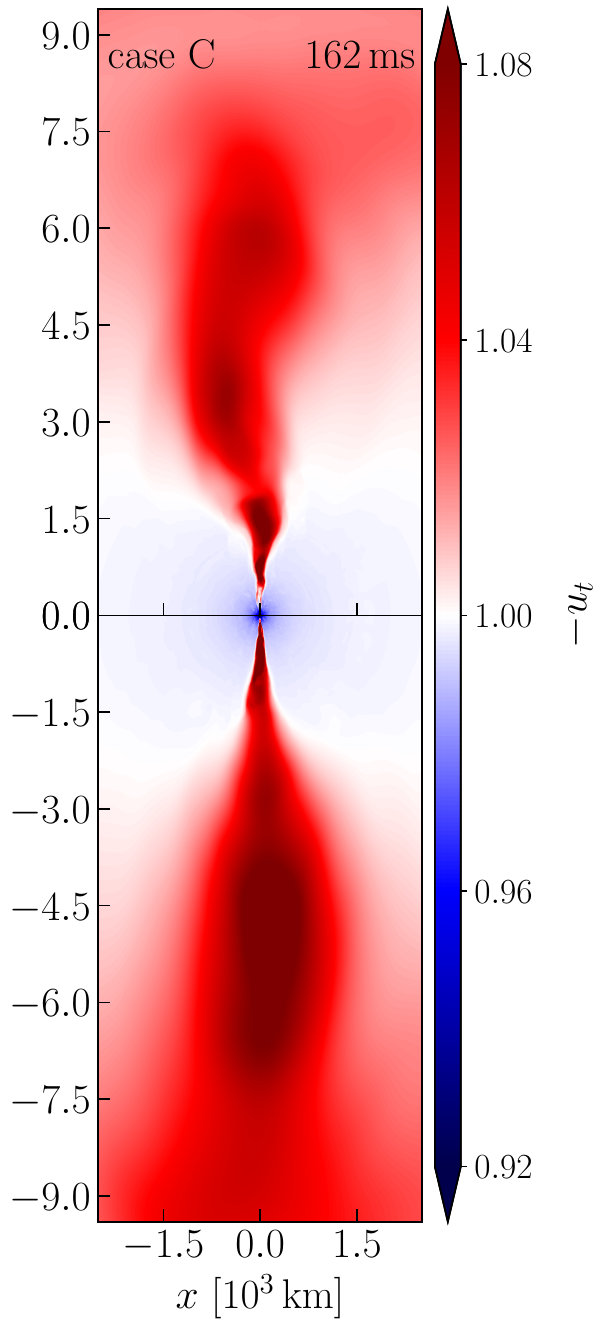} 
\caption{ Meridional view of $-u_t$ at 75\,ms after merger for cases A (top-left; $z>0$), B (top-center; $z>0$), C (bottom-left; $z<0$), D (bottom-center; $z<0$), and at 162\,ms after merger for case C on larger spatial scales (right).}
\label{fig:ut}
\end{center}
\end{figure}

Figure~\ref{fig:ut} shows unbound regions ($-u_t\!>\!1$) for all cases at $t\!=\!75$\,ms, and also at $t\!\simeq\!160$\,ms for case C. 
In cases A and C, we find rapid jet acceleration up to $\sim\!1500$\,km at $t\!=\!75$\,ms, reaching peak radial velocities of $v^r\!\simeq\!0.77$\,c and $0.69$\,c, respectively. 
Comparing the above velocities, we find that higher resolution favors a more efficient jet acceleration, although the qualitative dynamics remains the same.
In both cases, the jet  front extends to $\sim\!3\!\times\!10^3$\,km, but its velocity is slowed down to $\sim\!0.6$\,c by the dense pre-collapse outflow.
This stalling reflects the fact that the jet is still making its way across the outer part of the pre-collapse polar outflow.

\begin{figure}[t] 
\begin{center}
\includegraphics[width=1\columnwidth]{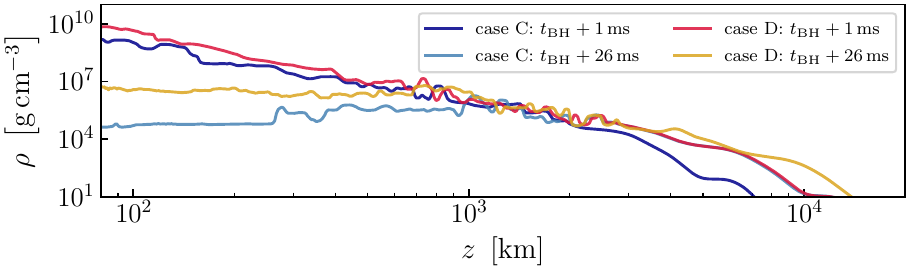}  
\caption{Rest-mass density profiles along the z-axis (south side) for cases C and D, at 1\,ms or 26\,ms after BH formation.}
\label{fig:rho_C_D}
\end{center}
\end{figure}

In case D, delaying the collapse by $\simeq\!25$\,ms is sufficient to strongly increase the level of baryon pollution due to the MNS-driven outflow, which results more massive and extended (Fig.~\ref{fig:rho_C_D}).
As a consequence, the incipient jet has to spend more time and energy to advance through the surroundings (e.g., compare, at the same time after collapse, case D in Fig.~\ref{fig:ut} and case C in Fig.~\ref{fig:utes}; see also Appendix \ref{colltime}). Moreover, it retains orders of magnitude higher density (Fig.~\ref{fig:rho_C_D}) and, in turn, lower terminal Lorentz factor.
These findings show that the collapse time, by strongly affecting the properties of the environment through which the incipient jet has to propagate, plays a fundamental role in determining the final jet energetics and breakout timescale. 

Finally, in case B, the unbound polar outflow produced by the MNS remains rather slow, with $v^r\lesssim\!0.3$\,c.
Moreover, its maximum terminal Lorentz factor is orders of magnitude lower compared to the BH-powered incipient jets of cases A, C, and D (see also Fig.~\ref{fig:ut} and Appendix \ref{colltime}).

Continuing the evolution of case C from 75 to 160\,ms after merger (Fig.~\ref{fig:ut}, rightmost panels), the maximum velocity starts growing again, reaching $\simeq\!0.74\,c$, while the fastest part of the incipient jet advances further before stalling again towards the end. At final time, the jet front extends beyond $\sim\!10^4$\,km. 
Overall, we observe a jet piercing through the environment in a series of finite steps, rather than experiencing a continuous advancement and acceleration. 
Only after breaking out of the dense material constituting the pre-collapse polar outflows, it will be able to fully accelerate, potentially reaching the high Lorentz factors required to power a GRB.
As a note of caution, at this stage we cannot establish which $\Gamma$ values can actually be achieved after the above breakout. 
If $\Gamma_\infty$ at the jet base maintains a high value, the reduced energy dissipation along the way could allow the jet to accelerate up to $\Gamma\!\sim\!100$, but this could only be proven by evolving the system for a longer time.  
We also note that moving to larger distances, the decrease in resolution could eventually impact the acceleration in a non-negligible way, causing the maximum velocity of the jet to be reduced by numerical dissipation. Consequently, our simulations can only demonstrate GRB-like terminal Lorentz factors close to the jet base, but not at larger distances.
%\mod{Consequently, GRB-like terminal Lorentz factors cannot be demonstrated at larger distances, but only close to the jet base.}

%

As final remark, we note that rest-mass densities within the incipient jet in cases A, C and D drop below $\!\rho^*\!=\!6.17\times10^{4}$\,g/cm$^3$, implying that a uniform density floor with $\rho_\mathrm{atm}\!=\!\rho^*$ would have affected the evolution. Going to larger time and spatial scales, after the jet breaks out of the surrounding environment and starts to accelerate to high Lorentz factors, the limitations of a high atmospheric floor would become prohibitive. In order to describe jets that are physically consistent and, at the same time, able to evolve into high-$\Gamma$ outflows compatible with GRBs, it is mandatory that their rest-mass densities remain always above the floor. 
Here, we have successfully conducted simulations where this condition is strictly maintained at all times.

% =====================
% Conclusion
% =====================
\section{Conclusions}

In this work, we have carried out simulations of BNS mergers with the GRMHD code \texttt{Spritz}, considering cases where the resulting MNS collapses into a BH about 25\,ms or 50\,ms after merger, plus one case with no collapse.
In the collapsing cases, the system naturally produces two qualitatively distinct, sequential outflows: a pre-collapse, magnetically driven polar outflow from the MNS and a post-collapse incipient jet powered by the accreting BH. 
The interaction between these two outflow components, leading to shock heating and potentially responsible for specific EM signatures, is self-consistently captured and discussed for the first time. 

All the emerging incipient jets show at their base terminal Lorentz factors and other properties that are in principle compatible with a GRB. Nonetheless, changing the collapse time from $\simeq\!25$\,ms to $\simeq\!50$\,ms has a major impact on the jet's ability to pierce through the environment, the latter being significantly more massive and extended in the second case.
We conclude that the collapse time is a key parameter to determine the final properties of the escaping jet, as well as the timescale to break out of the dense MNS-driven outflow.
Such a result offers a promising coherent framework for interpreting jet diversity in BNS mergers: 
very different outcomes, from fully choked outflows to successful GRB-like jets, could be explained by the ample range of possible MNS lifetimes before a BH-driven incipient jet is launched.

In our longest simulation (case C), which extends to $\simeq\!135$\,ms post-collapse, the BH-powered jet gets close to breaking out of the pre-collapse outflow, with its front reaching spatial scales of $\sim\!10^4$\,km.
The jet advancement through the surroundings is not continuous, but characterized by stalling episodes. 
This results from the combination of a discontinuous energy injection from the BH-disk engine and the interaction with a non-uniform and evolving environment.
Such an intermittent behaviour will likely be imprinted in the final GRB signal, and it may represent a dominant factor governing its variability.

In the no-collapse model, the MNS drives a steady, collimated outflow that remains much denser and slower than the BH-driven jets found in the other cases.
This result suggests that powering a GRB in absence of a BH may be challenging. Nonetheless, given the limited timescale and resolution of our simulations, such a scenario cannot be excluded.

A key technical advance in our investigation is the use of an unprecedentedly low density floor (radially declining as $\rho_{\rm atm}\!\propto \!r^{-6}$), which prevents artificial effects on all physically relevant outflows (dynamical ejecta, pre-collapse MNS wind, and BH-powered incipient jet) over the simulated spatial and temporal scales. 
For the ultimate goal of connecting the jet launching dynamics and early propagation to the final EM signatures and thus the GRB phenomenology, the absence of floor effects is a necessary condition.

Different caveats still affect our simulations, including the limited resolution (and consequent need for unrealistically strong pre-merger magnetic fields) and the use of a simple piecewise-polytropic EOS without neutrino transport. 
While overcoming such limitations is a central goal of our future work, this will not affect the main conclusions on the two distinct (MNS- and BH-driven) outflows, the role of the collapse time, or the advantage of a BH central engine in powering GRB jets.

% =====================
% Acknowledgements
% =====================
\vspace{0.5cm}
\begin{acknowledgments}

The authors thank Wolfgang Kastaun, Michail Chabanov, Lorenzo Ennoggi, and Liwei Ji for insightful discussions. 
The authors gratefully acknowledge the National Science Foundation (NSF) for financial support from grants AST-2009330, AST-2319326, PHY-2110338, PHY-2409706, OAC-2004044, OAC-2411068 as well as the National Aeronautics and Space Administration (NASA) for financial support from TCAN Grants No.~80NSSC18K1488, 80NSSC24K0100,  
and the Italian Ministry of Foreign Affairs and International Cooperation (MAECI; Grant No.~US23GR08). 
RC, BG, and AP acknowledge further support by the European Union under NextGenerationEU, Mission 4, Component 1, via the PRIN 2022 Project ``EMERGE'', Prot.~2022KX2Z3B (CUP C53D23001150006).
This research used resources from the Texas Advanced Computing Center's (TACC) Frontera and Vista supercomputer allocations (awards AST20021 and PHY20010). Additional resources were provided by the BlueSky, Green Prairies, and Lagoon clusters of the Rochester Institute of Technology (RIT) acquired with NSF grants PHY-2018420, PHY-0722703, PHY-1229173, and PHY-1726215.

\end{acknowledgments}

% =====================
% Appendix
% =====================
\appendix

\section{Collapse time and jet launching}
\label{colltime}

\begin{figure}
\begin{center}
\includegraphics[width=1\columnwidth]{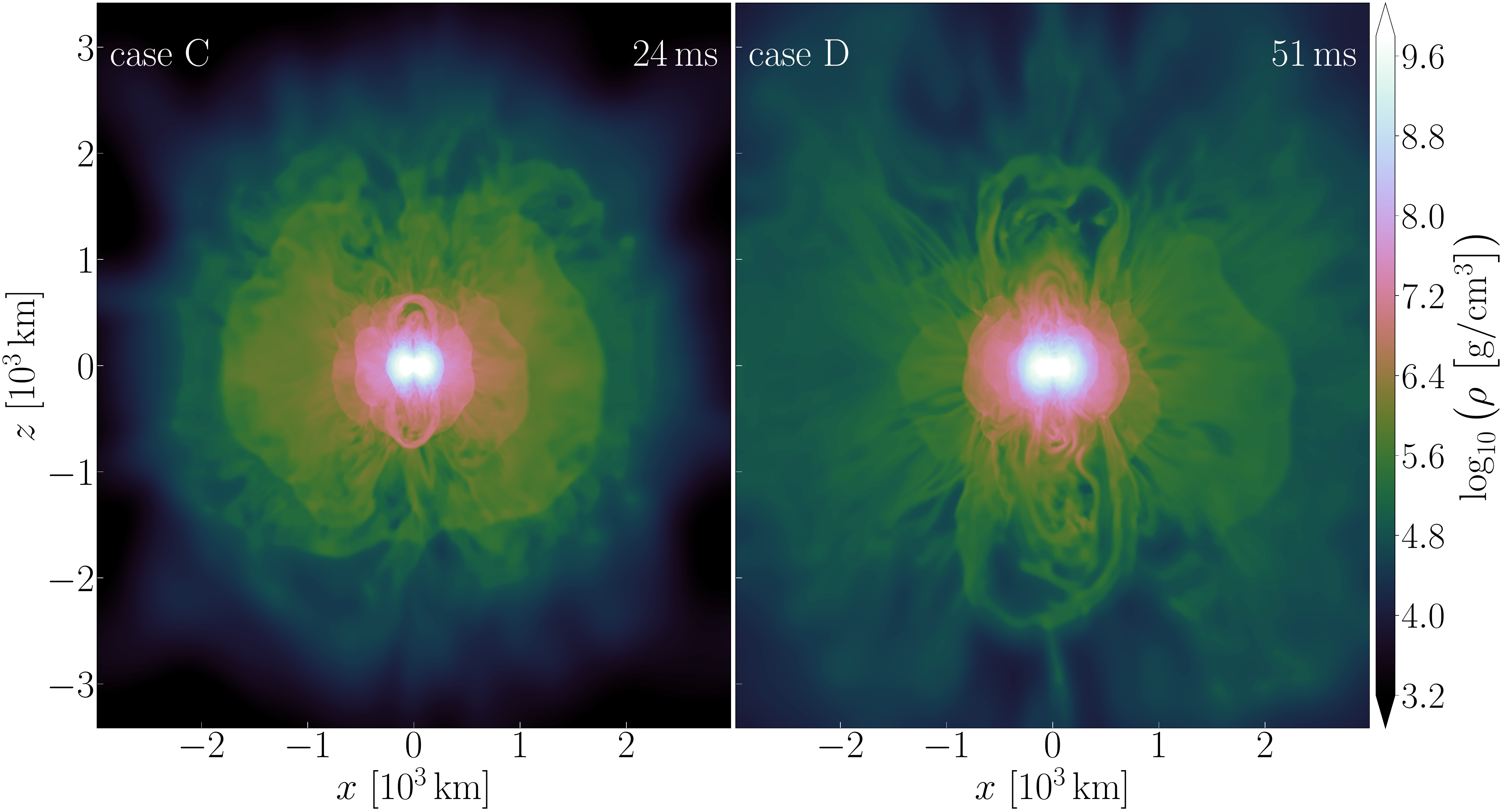}  
\caption{Snapshots of rest-mass density on the $xz$-plane around to the collapse time of the MNS remnant, i.e.~at $24$ and $51$\,ms after merger for cases C (left) and D (right), respectively.}
\label{fig:rhoCD}
\end{center}
\end{figure}

The survival time of the MNS remnant determines the duration of the pre-collapse polar outflow and the level of baryon loading in the funnel along the spin axis of the newly-formed BH. 
Figure~\ref{fig:rhoCD} shows a meridional view of rest-mass density around collapse time for cases C and D (at $24$ and $51$\,ms after merger, respectively).
A delay of only $\!\approx\!25$\,ms already produces very different conditions around the merger site. In case C, nascent polar outflows have just started to emerge and extend to $\sim\!500$--$700$\,km, while in case D the same outflows have expanded to $\sim\!2000$\,km and dominate the mass loading along the $z$-axis. In the latter case, while the outflow front is expanding, we see that a magnetized, overpressured cocoon is inflated, resulting in a more difficult propagation environment for the BH-driven incipient jet launched afterwards.
\begin{figure}[t!]
\begin{center}
\includegraphics[width=1\columnwidth]{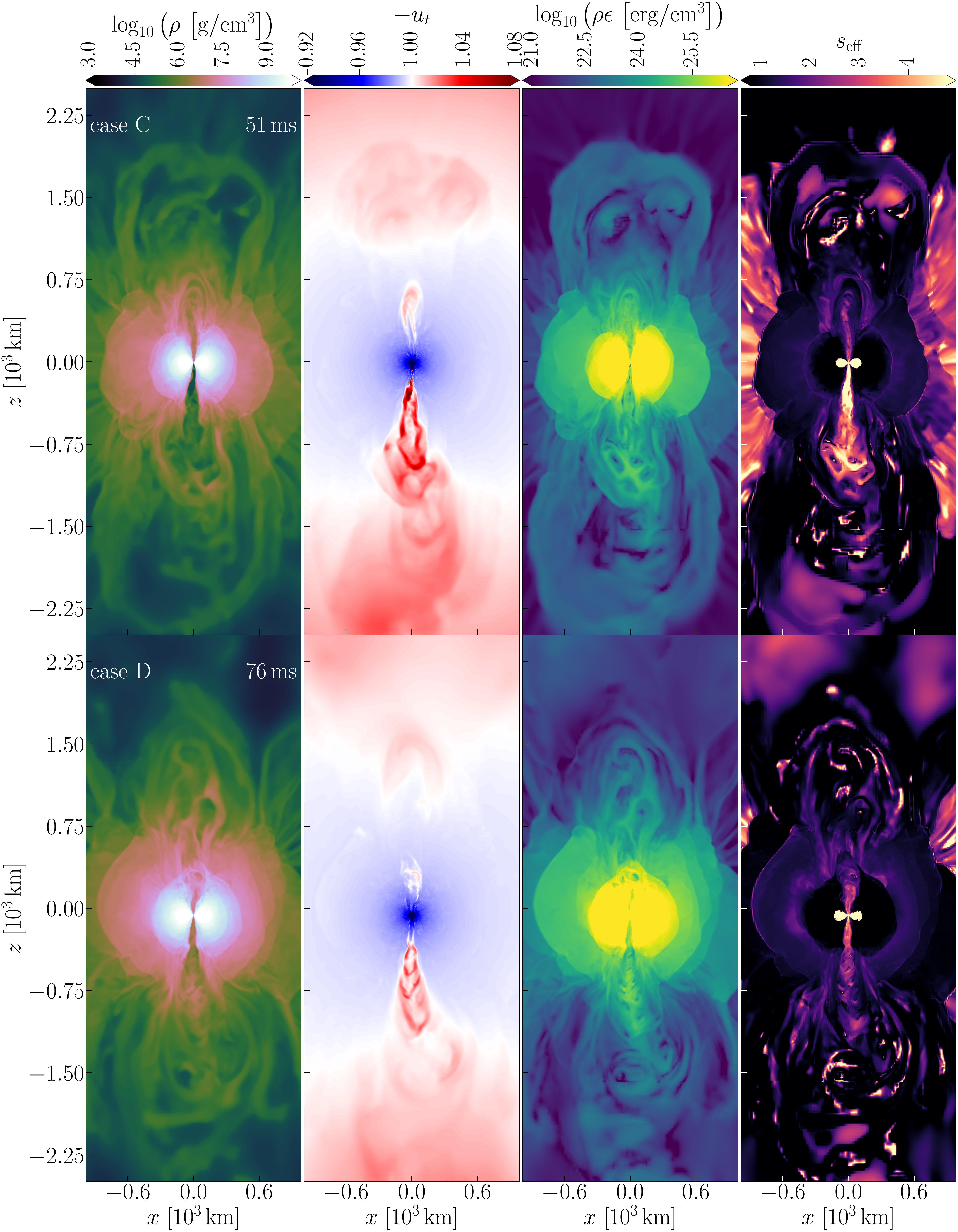}  
\caption{Meridional view of the rest-mass density, $-u_t$, internal energy density $\rho\epsilon$ and effective specific entropy $s_{\rm eff}$ for cases C (top panels) and D (bottom panels) at $\approx\!25$\,ms after BH formation.}
\label{fig:jetCD}
\end{center}
\end{figure}

After collapse, the emerging jet encounters the MNS-driven polar outflow and makes an effort to  drill through and eventually break out. 
Figure~\ref{fig:jetCD} compares this interaction 
$\approx\!25$\,ms after BH formation for cases C (top) and D (bottom) in terms of rest-mass density $\rho$, $-u_t$ (a proxy for unbound matter), internal energy density $\rho\epsilon$, and effective specific entropy $s_{\rm eff}\!\equiv\! \ln(P/ \rho^{\gamma_{\rm eff}})$, where $\gamma_{\rm eff}\!\equiv\!P/(\rho \epsilon) + 1$ (indicative of shock heating). 

In both cases, the jet emerges earlier on the south side ($z\!<\!0$). 
This asymmetry likely results from a north/south difference in the pre-collapse outflow morphology, common to the two cases C and D. Focusing on the south side, we see how the denser funnel in case D forces the jet to spend more energy and time to propagate, implying significantly different breakout timescale and final properties of the escaping jet.

As already emphasized in the main text, jet's energy dissipation is accompanied by shock heating, well visible in $\rho\epsilon$ panels. For case C, south side, this effect is particularly evident also looking at $s_{\rm eff}$. 
We note that these shocks are strongly localized.
In the angular direction, they are limited within half-opening angles of 10-15 degrees from the jet propagation axis. In the radial direction, they trace the relatively slow advancement of the BH-driven jet front, where we have the sharpest transition from faster, highly unbound, low density material to the outer much slower and denser environment. 
The corresponding local peaks in internal energy density and effective specific entropy can reach much larger values than those observed at the outer front of the expanding MNS-driven outflows (see in particular case C, south side in Figure~\ref{fig:jetCD}).
Moreover, we expect this energy dissipation to be most prominent while the jet is still piercing through the dense MNS-driven material, i.e.~within a small time window from a few tens of ms after collapse until the eventual breakout of the jet into the outer dynamical ejecta.
Due to such distinctive properties, this shock heating may leave a unique signature in the EM emission, possibly producing a precursor signal to the GRB or directly contributing to the latter.

\begin{figure}
\begin{center}
\includegraphics[width=1\columnwidth]{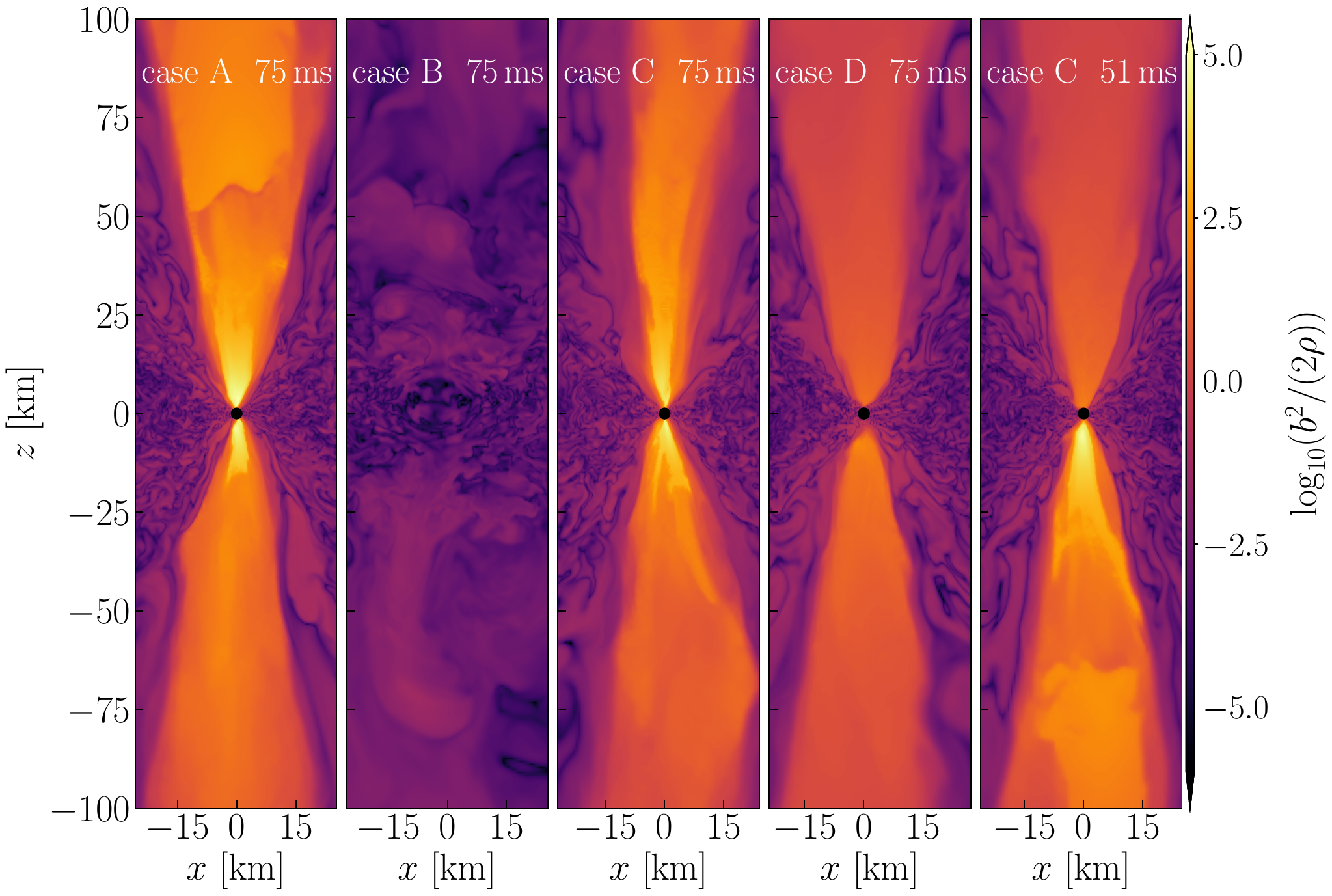}  
\caption{Magnetization $b^2/(2\rho)$ on xz-plane for cases A-D at 75\,ms after merger (first four panels), and for case C at $\approx\!50$\,ms after merger (last panel). }
\label{fig:magnetization}
\end{center}
\end{figure}

We next examine the magnetization near the BH or MNS remnant as a proxy for the acceleration potential of the incipient jet. 
In particular, Figure~\ref{fig:magnetization} shows the magnetic-to-rest-mass energy density ratio $b^2/(2\rho)$ for cases A--D at 75\,ms after merger and, only for case C, also at about 50\,ms after merger. When this quantity is much larger than unity, it corresponds approximately to the terminal Lorentz factor $\Gamma_\infty$. 

Close to the BH, cases A and C reach values above $10^{4}$. Comparing cases C and D at about $25$\,ms after collapse (two rightmost panels in Fig.~\ref{fig:magnetization}), case C exhibits orders of magnitude higher magnetizations on the south side, consistent with a cleaner funnel and an incipient jet of smaller density (see Figure~\ref{fig:jetCD}). Finally, we also note that the non-collapsing case B does not show a magnetization capable of producing significant acceleration, suggesting that the polar outflow will remain rather slow also at increasingly large distances.

Overall, the comparison between cases C and D shows the strong impact of the collapse time in the context of jet launching.
As concluded in the main text, we find that an earlier collapse limits baryon pollution in the funnel and favors a faster jet propagation which is also energetically less expensive; later collapse allows the pre-collapse MNS-driven outflow to inflate a heavier and more extended cocoon, delaying the jet breakout and dimming the EM emission powered by the jet itself.
A denser and more massive cocoon should also enhance photospheric emission and delay high-energy transparency, potentially offering multiwavelength diagnostics of the collapse time.

%%%%%%%%%%%%%%
\section{Numerical floor prescription}
\label{floorpres}

As described in the main text, in all our simulations we employ a radially decreasing numerical rest-mass density floor based on the power-law profile $\rho_\mathrm{atm}\!=\!\rho^*\,(r/r^*)^{-6}$
(with $\rho^*\!=\!6.17\times10^{4}$\,g/cm$^3$ and $r^*\!=\!74$\,km). The sixth power of the radial distance, shown in \cite{Pavan2021,Pavan2023} to be a good choice to avoid artificial effects at increasing large scales, corresponds to the lowest numerical floor ever employed in BNS merger simulations.
As already noted, radially decreasing profiles where also employed in previous work. For instance, in \cite{hayashi2024jet}, the floor scales as $\rho_\mathrm{atm}\!\propto\!r^{-3}$ down to 0.166\,g/cm$^3$ and then remains constant, giving a floor level higher than ours above $\sim\!700$\,km distance and, e.g., $\sim\!1000$ times larger at 2000\,km.
\begin{figure}
\begin{center}
  \includegraphics[width=0.9\columnwidth]{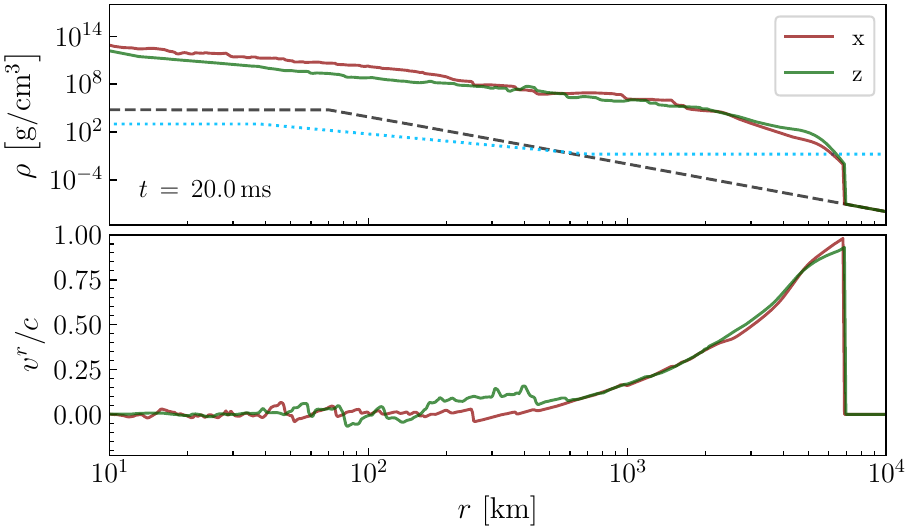}  
  \caption{Rest-mass density (top) and radial velocity (bottom) along the positive $x$ and $z$ axes at 20\,ms after merger, for case A. Rest-mass density floor profiles in the top panel are shown as a black dashed line for our simulations and a light blue dotted line for \cite{hayashi2024jet} (see text).}
  \label{fig:rhovr}
\end{center}  
\end{figure}

In Figure~\ref{fig:rhovr}, top panel, we show the rest-mass density profiles along $x$- and $z$-axis obtained at 20\,ms after merger in case A, compared to the floor employed in our work and in \cite{hayashi2024jet}.
From such a comparison, we see that the densities of the dynamical ejecta produced during the merging process remain orders of magnitude above our floor, allowing us to correctly describe them up to their natural front with velocities of about $0.9\,c$ (Figure~\ref{fig:rhovr}, bottom panel). Coversely, employing the same floor as in \cite{hayashi2024jet} would have started affecting such a high-velocity tail already at 20\,ms after merger, eventually becoming a critical limitation at increasingly large distances.

\begin{figure}
\begin{center}
  \includegraphics[width=1\columnwidth]{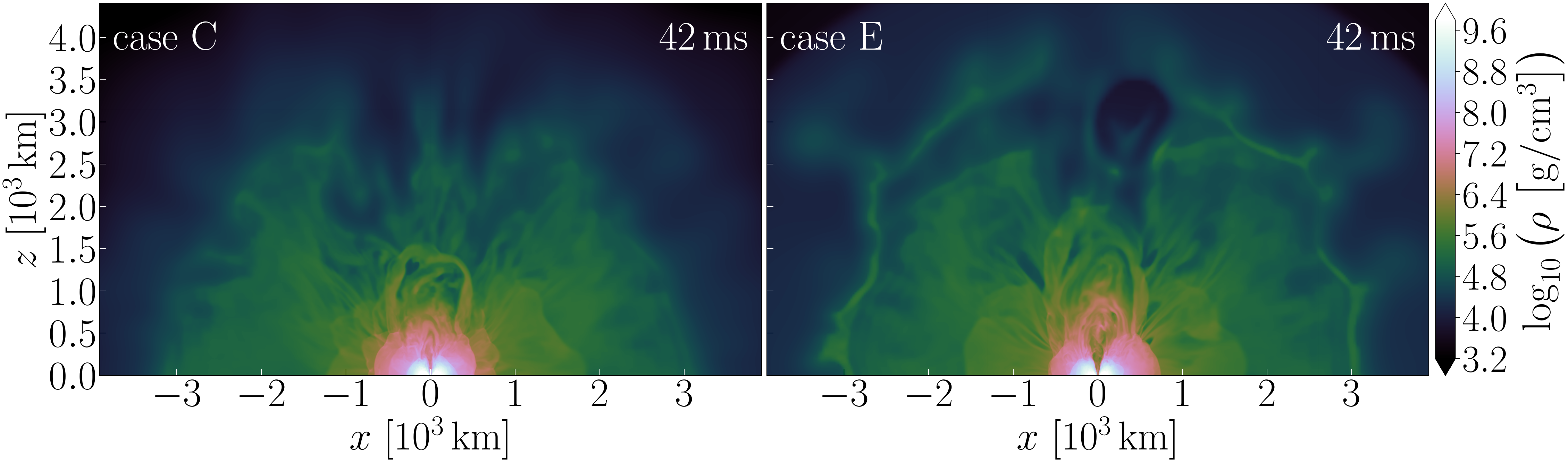}  
  \caption{Meridional view (northern hemisphere only, $z>0$) of the rest-mass density at 42\,ms after merger, for case C (left panel) and case E (right panel).}
  \label{fig:atmograd}
\end{center}  
\end{figure}

To further assess the effects of the floor, we also performed an additional simulation (case E) analogous to case C but employing a rest-mass density floor profile as $\rho_{\rm atm}\!\propto\! r^{-2}$. 
In this case, we only evolved up to 42\,ms after merger, i.e.~less than 20\,ms after the time of collapse to a BH.
A meridional view of the rest-mass density comparing cases C and E at such time is shown in Figure~\ref{fig:atmograd}.
At this early evolution stage, we can already observe relevant differences in both the dynamical ejecta and the post-merger outflows.
Going to later times, such discrepancies will inevitably grow more and more.

The above results indicate that a floor with $\rho_{\rm atm}\!\propto\!r^{-2}$ is not sufficiently low for a reliable description of the emerging incipient jet and its propagation across the environment. 
\begin{figure}[t]
\begin{center}
\includegraphics[width=0.98\columnwidth]{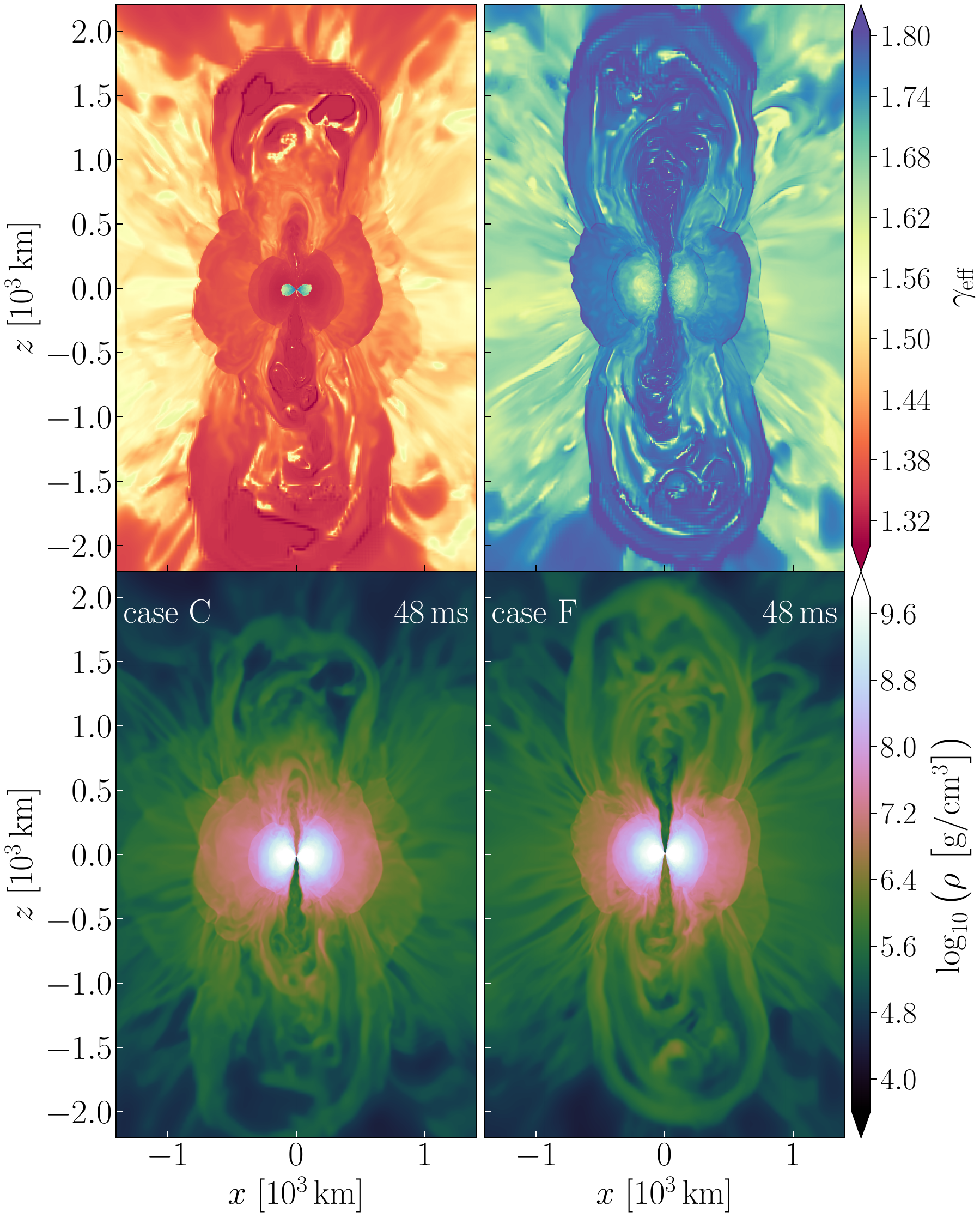}  
\caption{Meridional view of the effective adiabatic index $\gamma_\mathrm{eff}$ (top panels; see text for definition) and rest-mass density (bottom panels) at 48\,ms after merger for cases C (left) and F (right).}
\label{fig:gammaeff}
\end{center}
\end{figure}

%%%%%%%%%%%%%%%%%
\section{Step function for $\gamma_{\rm th}$}
\label{stepfunc}

The simulations presented in this work employ a step function for the adiabatic index $\gamma_{\rm th}$ of the thermal component of the hybrid EOS. 
Specifically, for rest-mass densities above $10^{10}\,\text{g/cm}^3$ we set $\gamma_{\rm th}\!=\!1.8$, which is a typical value adopted in the literature, while below such densities we impose $\gamma_{\rm th}\!=\!4/3$, corresponding to the ultra-relativistic limit. 
The latter choice reflects the expectation that, in realistic microphysical EOS, the effective ideal-gas adiabatic index \footnote{This corresponds to the adiabatic index that a simple ideal gas would have for given values of $\rho$, $P$, and $\epsilon$.} $\gamma_{\rm eff}\!\equiv\!P/(\rho \epsilon) + 1$ becomes 4/3 at low rest-mass densities, as discussed in \cite{Bauswein2010}.
The choice of $10^{10}\,\text{g/cm}^3$ as threshold value is also consistent with the results of \cite{Bauswein2010} (see Figure~2 therein). 

In Figure~\ref{fig:gammaeff} (top left panel), we present a meridional view of $\gamma_{\rm eff}$ for case C at 48\,ms after merger.
In the lower left panel, we show instead the rest-mass density.
For comparison, we performed an additional simulation (case F) in which we maintained the same adiabatic index $\gamma_{\rm th}\!=\!1.8$ at any density. Case F is also reported in Figure~\ref{fig:gammaeff} (right panels). 

As shown in the upper panels, the value of $\gamma_{\rm th}$ at low densities (i.e.~$\rho\!<\!10^{10}\,\text{g/cm}^3$) essentially determines the EOS behavior in terms of $\gamma_{\rm eff}$ for both the emerging incipient jet and the environment in front of it. 
In those regions, $\gamma_{\rm eff}$ is indeed close to 4/3 in case C, and around 1.8 in case F. 
Comparing the rest-mass density in the lower panels, we see that the different choice of the adiabatic index at low densities has a strong impact on the system dynamics. In particular, the incipient jet emergence is favored in case F. 

In conclusion, (i) introducing the step function for $\gamma_{\rm th}$ allows us to reproduce a realistic adiabatic index of $\approx\!4/3$ for the jet and its propagation environment, while maintaining the same EOS behavior at $\rho\!>\!10^{10}\,\text{g/cm}^3$, and (ii) the resulting evolution is rather different from a case where $\gamma_{\rm th}$ is instead constant and equal to 1.8.  

%%%%%%%%%%%%%%%%%%

% =====================
% Bibliography
% =====================

\bibliography{references}
\bibliographystyle{aasjournalv7}

\end{document}